\documentclass[pre,superscriptaddress,twocolumn,footinbib]{revtex4-1}
\pdfoutput=1
\usepackage[colorlinks=true,urlcolor=blue]{hyperref}
\usepackage{amsfonts}
\usepackage{graphicx}
\usepackage{placeins}
\usepackage{amsmath}
\usepackage{xcolor}
\usepackage{color}
\usepackage{bm}

\definecolor{red}{rgb}{0.75,0,0}
\definecolor{blue}{rgb}{0,0,0.75}
\definecolor{green}{rgb}{0,0.5,0}

\begin{document}

\title{Cellular geometry controls the efficiency of motile sperm aggregates}

\author{D. J. G. Pearce}
\affiliation{Instituut-Lorentz, Universiteit Leiden, P.O. Box 9506, 2300 RA Leiden, The Netherlands}
\author{L. A. Hoogerbrugge}
\affiliation{Kavli Institute of Nanoscience, Delft University of Technology, Delft, The Netherlands}
\affiliation{Instituut-Lorentz, Universiteit Leiden, P.O. Box 9506, 2300 RA Leiden, The Netherlands}
\author{K. A. Hook}
\affiliation{Department of Biology University of Maryland College Park, MD 20742, U.S.A.}
\author{H. S. Fisher}
\affiliation{Department of Biology University of Maryland College Park, MD 20742, U.S.A.}
\author{L. Giomi}
\affiliation{Instituut-Lorentz, Universiteit Leiden, P.O. Box 9506, 2300 RA Leiden, The Netherlands}

\begin{abstract}
Teams of cooperating sperm have been found across several vertebrate and invertebrate species, ranging from sperm pairs to massive aggregates containing hundreds of cells. Although the biochemical mechanisms involved in the aggregation process are still unclear, it was found that aggregation can enhance the mobility of the cells, thus offering an advantage during fertilization. Here, we report a thorough computational investigation on the role of cellular geometry in the performance of sperm aggregates. The sperm head is modeled as a persistent random walker characterized by a non-trivial three-dimensional shape and equipped with an adhesive region, where cell-cell binding occurs. By considering both a simple parametric head shape and a computer reconstruction of a real head shape based on morphometric data, we demonstrate that the geometry of the head and the structure of the adhesive region crucially affects both the stability and mobility of the aggregates. Our analysis further suggests that the apical hook commonly found in the sperm of muroid rodents, might serve to shield portions of the adhesive region and promote efficient alignment of the velocities of the interacting cells.
\end{abstract}

\maketitle

\section{Introduction}

In most species, the number of sperm available for fertilization far outnumber ova, which gives rise to fierce competition at the cellular scale \cite{Parker:1970}. 
The highly efficient and streamlined shape of most sperm cells is the product of intense selective pressure to improve swimming performance. In a rare number of systems, sperm form cooperative groups, or aggregates \cite{Moore:2002,Higginson:2010,Pizzarri:2008,Immler:2008};
this unique behavior is believed to improve the swimming performance of the cells involved, compared to individual cells, and therefore the chances of successful fertilization \cite{Immler:2008}. Yet, {\em in vitro} studies have shown inconsistent results \cite{Pizzarri:2008} and the underlying physics of the associations remains elusive \cite{Yang:2008,Tung:2017,Schoeller:2018}.

Among mammals, the natural variation observed within muriod rodents offers important insight into how cell shape and orientation can mediate collective motion of sperm \cite{Breed:2005}. For example, the sperm of house mice ({\em Mus musculus})
form groups in which the head of a sperm is bound to either the head or the tail of another cell, whereas in the Norway rat ({\em Rattus norvegicus}), sperm form comet-like aggregates in which all the cells are bound at the head and preserve the head-tail directionality of a single cell \cite{Immler:2007}. These morphological variations result in a substantial difference in the aggregate swimming performance: while in the Norway rat, sperm groups swim faster than single cells, no speed advantage is found in the house mouse. Train-like aggregates are also found in the wood mouse ({\em Apodemus sylvaticus}), where aggregation is known to improve the swimming performance \cite{Moore:2002}. In most muroid rodents, sperm have a falciform head with an apical hook that is thought to facilitate the formation \cite{Moore:2002} and/or stabilization \cite{Immler:2007} of aggregations, although other scenarios have also been suggested \cite{Firman:2009}. Although the molecular mechanisms that regulate sperm-sperm adhesion in these systems are not well understood, in house mice, transmembrane glycoproteins more typically associated with sperm-egg binding are likely involved \cite{Han:2010}.

In previous work, we studied sperm from {\em Peromyscus} rodents and analyzed how the swimming velocity of the aggregates is modulated by group size \cite{Fisher:2014}. By combining fine-scale imaging of living cells and a simple two-dimensional mechanical model of sperm aggregates (inspired by the Vicsek model \cite{Vicsek:1995} and the statistical mechanics approach to collective behavior \cite{Czirok:1996,Couzin:2002,Buhl:2006,Ballerini:2008,Peruani:2012,Vicsek:2012}) we demonstrated that the average velocity of an aggregate does indeed increase with its size as the group moves more persistently. This benefit, however, is offset in larger aggregates as the geometry of the group forces sperm to swim against one another. This result is a non-monotonic relationship between aggregate size and average velocity with an optimum that is predominantly determined by the geometry of the sperm head \cite{Fisher:2014}. The underlying mechanism leading to this optimality is straighforward: by forming tightly packed head-to-head aggregates, sperm are able to ease each other’s directional fluctuations, resulting in a straighter trajectory, thus a greater average velocity. Large aggregates, however, tends to be isotropic and this forces the sperm to swim against one another, thus reducing the aggregate speed. 

In this work, we report a thorough computational investigation on the role of cellular geometry in the performance of sperm aggregates, as well as on the structure of the adhesive region of the cell head. For this purpose, the sperm head is modeled as an ellipsoid of revolution. Although considerably less structured than the actual head shape found in most muroid rodents, this is a simple parametric shape with enough morphological features (e.g. slenderness, oblatness, etc.) and provides insight into the effect of the head shape on the spatial organization and the motility of the aggregates. Furthermore, with the help of a data-based computer reconstruction of a {\em Peromyscus maniculatus} sperm head, we explore the interplay between the geometry of the apical hook and the structure of the adhesive region, and identify a number of highly efficient configurations that could provide a basis for future experimental studies.  

\section{\label{sec:model}The model}

\begin{figure}[t]
\centering
\includegraphics[width=\columnwidth]{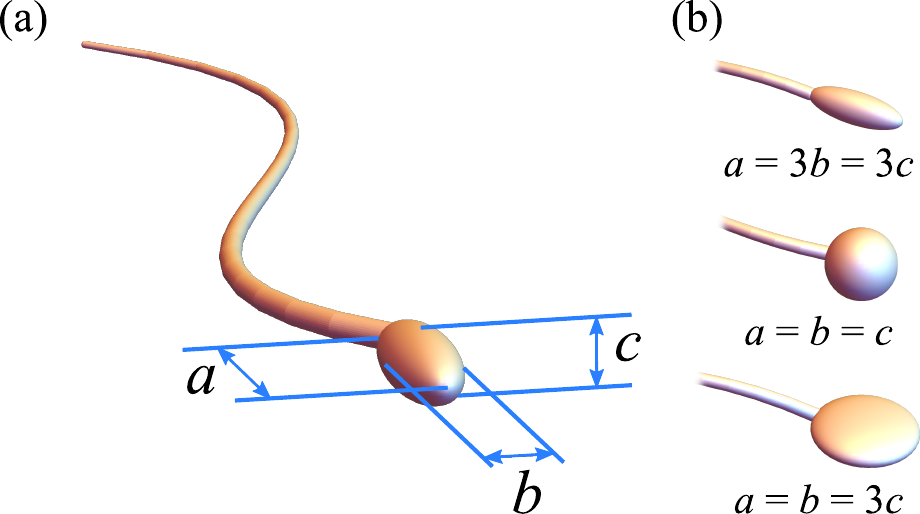}
\caption{\label{fig:sketch}(a) Schematic representation of the geometrical model used in the simulations. The cell's head is represented as an ellipsoid whose axes are $a$, $b$ and $c$.  (b) Upon adjusting the ratios $b/a$ and $c/a$, one can reproduce various head morphologies.}
\end{figure}

Our computational model aims to reproduce the physical interactions between sperm cells arising from steric repulsion and linkage through adhesive molecules. Each cell is modeled as a persistent random walker \cite{Fisher:2014}, subject to a constant propulsive force and a random torque and whose dynamics is governed by the over-damped Newton equations. When two sperm cells come into contact they experience a repulsive force sufficient to prevent them from overlapping. Cell-cell adhesion is modelled by a spring-like adhesion molecule. These adhesion molecules are localized to specific regions on the surface of the sperm head, and, if the adhesive regions of two neighbouring heads come into contact, a tether is formed between them. These tethers have a finite strength and will break if stretched too far. At each time step the cumulative force on each sperm from collisions, self propulsion and adhesion is calculated; this force is then used to update the position of each sperm cell.

The sperm head is modeled as a three-dimensional simplex: a set of triangles connected along the edges in such a way to form a closed polyhedron. This allows us to reproduce any desired morphological features including the apical hook found in the sperm cells of most muroid rodents \cite{Breed:2005}. The simplest non-trival three-dimensional shape suitable for a geometrical model of the sperm head is the ellipsoid of revolution (Fig. \ref{fig:sketch}a). Although less complex than the typical head shape found in rodents, this is a simple and yet sufficiently rich parametric shape, whose morphological feature can be entirely described in terms of the three parameters $a$, $b$ and $c$ (Fig. \ref{fig:sketch}a). Adjusting the ratio between these lengths permits us to reproduce various kinds of shapes (Fig. \ref{fig:sketch}b).

\section{\label{sec:results}Results}

\subsection{Aggregates travel faster than individual cells}

\begin{figure}[t]
\includegraphics[width = \columnwidth]{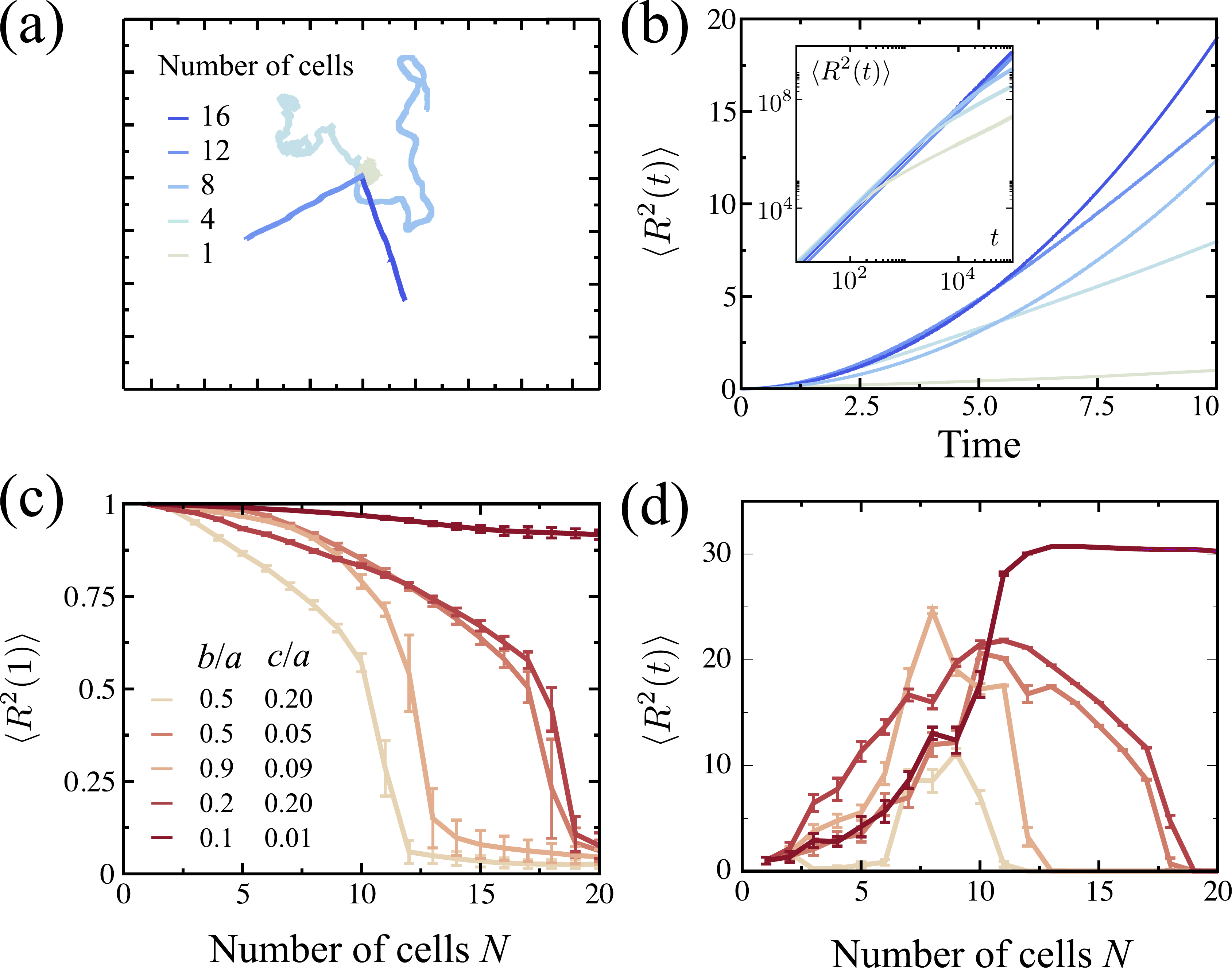}
\caption{\label{fig:N_scaling} (a) Aggregates are on a random walk, with no directed motion. The random walk appears to have a diffusion coefficient that depends on the number of sperms in the aggregate. (b) This can be quantified by examining the mean square displacement for an aggregate. This demonstrates a change from a ballistic to diffusive regime that depends on the number of sperms. (c) Forming an aggregate reduces the instantaneous movement speed of the sperm, this is most pronounced for wider cells. (c) The relative advantage of forming an aggregate depends on both the size of the aggregate and the geometry of the sperm head shape. For these simulations the adhesive region of the sperm cells is at the tip of the head.}
\end{figure}

\begin{figure*}
\includegraphics[width = \textwidth]{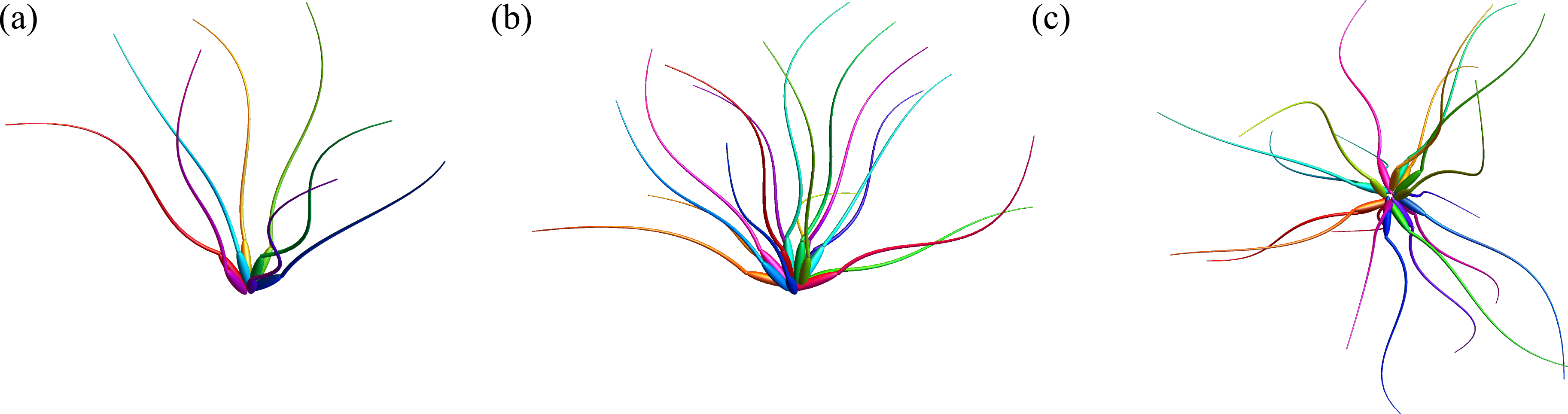}
\caption{\label{fig:3D_packing} Images showing how the typical configurations of sperm aggregates changes with number of cells. (a) Here $N=8$ cells retain a high degree of polarization and the packing reduces rotational diffussion. (b) For $N=16$ cells the total polarization is reduced, many cells on the periphery of the aggregate are pushing inwards rather than forwards. In this configuration the rotational diffusion is very low but the speed is significantly reduced. (c) $N=20$ cells arrange with no net polarization. The speed of the aggregate is now close to zero with no persistent direction. }
\end{figure*}

Each individual {\em in silico} sperm swims forward while undergoing rotational diffusion, mimicking the erratic swimming behaviour of real sperm. There is no preferred direction or target in the simulation, therefore sperm cells perform a persistent random walk (Fig. \ref{fig:N_scaling}a). The latter is characterized by a short-time balistic regime, in which the mean-square displacement scales quadratically with time (i.e. $\langle R^2(t) \rangle \sim t^2$, where the angular brakets indicate a statistical average), and a diffusive regime, in which $\langle R^2(t) \rangle \sim t$. The ballistic motion occurs over a time interval that is shorter than or comparable to the reorientation time of the sperm. The diffusive regime, on the other hand, is found when the time interval is greater than the reorientation time of the sperm (Fig. \ref{fig:N_scaling}b).

When many cells crowd together, the available space for each sperm head is restricted, leading to a reduction in the amount of rotational motion each sperm head experiences while swimming (Fig.~\ref{fig:N_scaling}a). This results in a longer reorientation time for a sperm aggregate, thus a larger persistence (Fig.~\ref{fig:N_scaling}b). As a consequence, the distance covered by a sperm aggregate within a given time interval is larger than in the case of individual cells, effectively making aggregates faster. Such a benefit, however, also depends on the traveling time. For instance, as shown in Fig.~\ref{fig:N_scaling}b, groups of $N=4$ sperm cells are faster than groups of $N=16$ only in short time periods, whereas this pattern reverses at longer durations.

Whereas the average velocity (i.e. $\sqrt{\langle R^{2}(t) \rangle}/ t$) increases with the size $N$ of the aggregates, their speed is a monotonically decreasing function of $N$. This originates from the fact that, due to their ellipsoidal shape, cells are always partially splayed toward the mean direction of motion of the aggregate. As a consequence, part of the propulsion provided by the swimming cells is lost, leading to a reduction in the instantaneous speed of the aggregate (Fig.~\ref{fig:N_scaling}c and Fig.~\ref{fig:3D_packing}). The instantaneous speed of all aggregates monotonically decreases with additional sperm cells, eventually reaching zero when groups have no net polarization and cannot move. This effect is far more pronounced in sperms with wider heads, because slender cells are able to pack more efficiently with nearly parallel orientations in far larger numbers.

To quantify the relative velocity increase caused by aggregation, we have compared the mean-squared displacement of aggregates of different size at a fixed time $t$ (Fig. \ref{fig:N_scaling}d). The latter was chosen so as to reflect the advantages of the rotational damping and the disadvantages of loss of instantaneous speed. We normalize mean-squared displacement of the aggregates by that of an individual cell to obtain a relative velocity increase. With this choice of $t$, the optimal number of cells ranges between 7 and 10, with a maximum velocity increase of nearly thirty times. This behavior exhibits  strong dependence on the geometry of the sperm cell head with slender cells generally seeing a far greater increase in distance travelled which is true even for very large aggregates. The origin of this property is two-fold: on the one hand, the instantaneous speed of the aggregate is higher for slender cells, as mentioned earlier. On the other hand, very prolate ellipsoid pack more efficiently than other shapes. This hinders the reorientation of the cells inside an aggregate, thus enhancing the rotational damping described earlier.

\subsection{Head geometry strongly affects the performance of the aggregate}

\begin{figure}[t]
\includegraphics[width = \columnwidth]{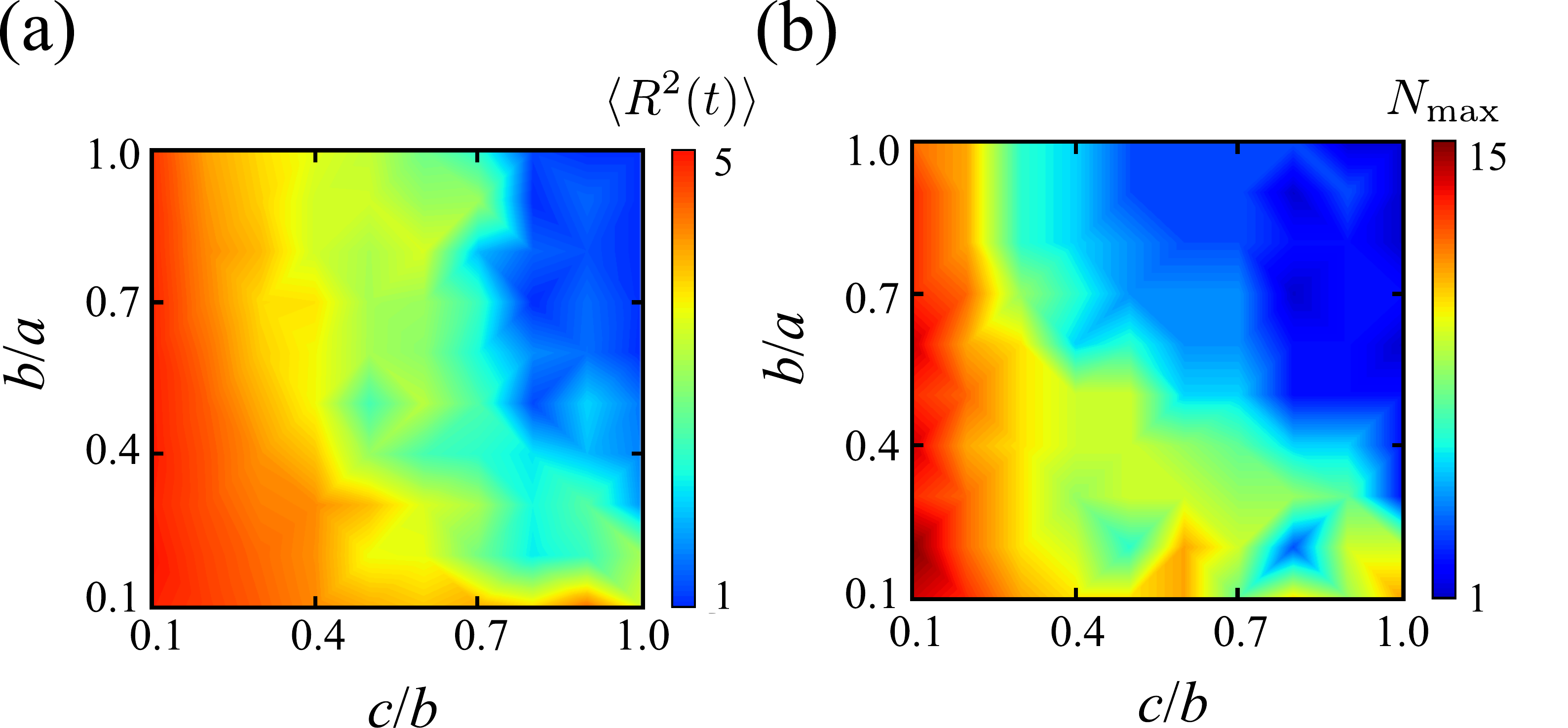}
\caption{\label{fig:GeoPhase} (a) Maximum speed increase of an aggregate of sperm cells with different shaped heads. (b) Optimum aggregate size for sperm cells with different head shapes. For these simulations the adhesive region of the sperm cells is at the tip of the head.}
\end{figure}

In order to shed light on the effect of sperm head geometry on the performance of motile aggregates, we performed simulations for a wide range of head aspect ratios ($b/a$ and $c/a$) and aggregate size ($N$). Fig.~\ref{fig:GeoPhase}a shows the maximum relative velocity of an aggregate for different aspect ratios. It is immediately clear that the fastest aggregates contain cells with a smaller aspect ratio, due to the fact that slender cells can form tightly packed aggregates where the velocities of the individual cells are roughly parallel. Interestingly, the strongest dependence observed here is on the second aspect ratio $c/b$, which implies that prolate (i.e. rod-like) and oblate (i.e. plate-like) heads have comparable efficiency upon aggregation.  Intuitively, this originates from the fact that plate-like sperm heads are able to stack, thus increasing the alignment of the velocities of individual cells. Furthermore, for a given volume, plate-like heads have significantly larger area than both spherical and rod-like heads, hence more space for membrane bound adhesion molecules \cite{Breed:2005}. 

Fig.~\ref{fig:GeoPhase}b shows the optimal size of sperm aggregate for different shaped sperm cells. Consistent with the previous considerations, flatter sperm heads obtain the greatest advantage when they form larger aggregates, this again is due to their ability to pack many cells effectively with highly aligned velocities. For very spherical sperm heads, there is no advantage to forming an aggregate, and we see that the optimal group size consists of a single cell (blue region in Fig.~\ref{fig:GeoPhase}b).

\subsection{The structure of the adhesive region strongly affects the performance of the aggregate}

\begin{figure}[t]
\includegraphics[width = \columnwidth]{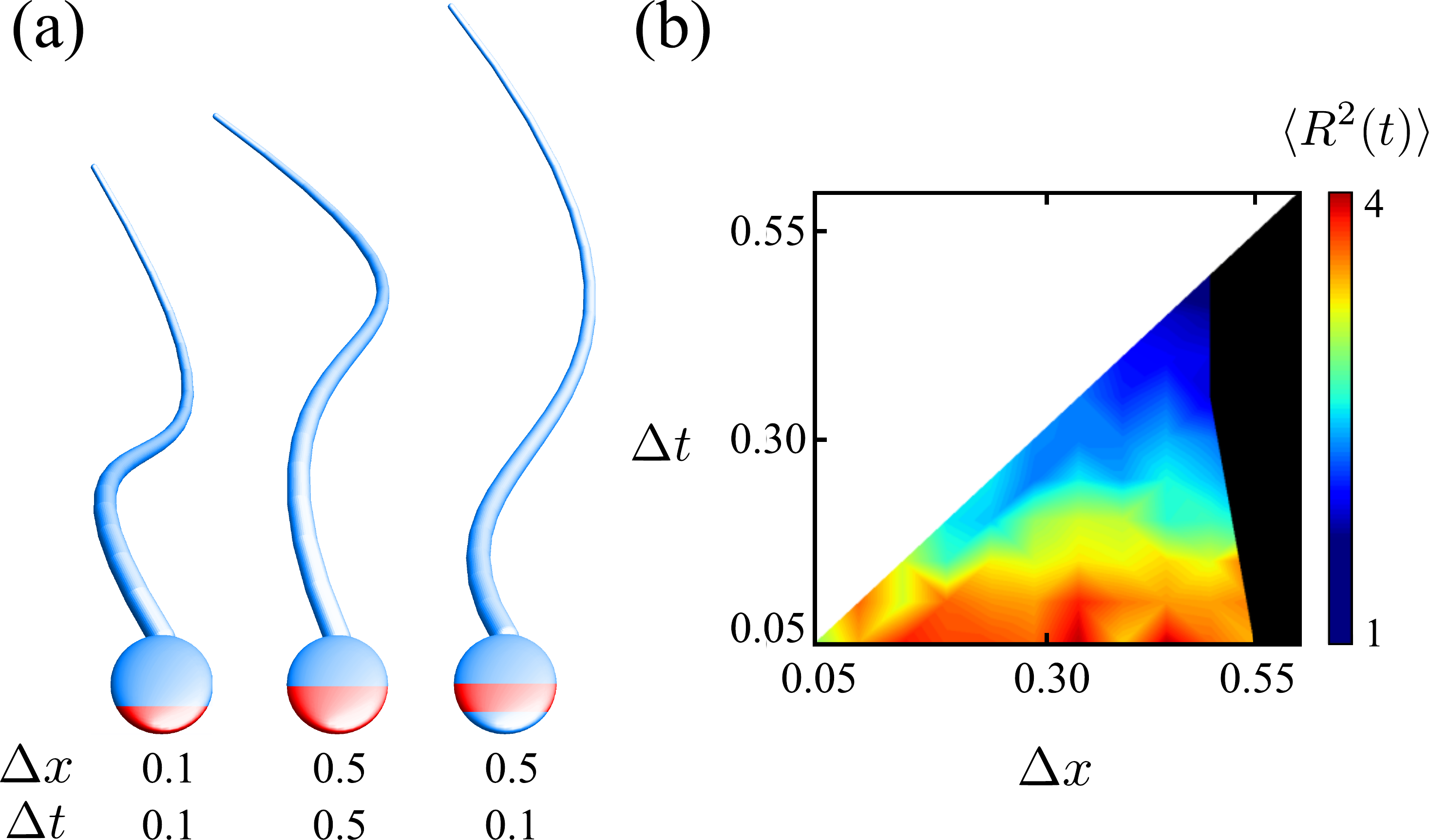}
\caption{\label{fig:AcroPhase} (a) The adhesive region of the sperm head can be parameterized by two lengths corresponding to the starting position $\Delta x$ and the thickness $\Delta t$. (b) Velocity increase obtained from the simulations of five different variants associated with different values of $\Delta x$ and $\Delta t$. The best performance is obtained when the adhesive strip occupies a small portion of the sperm head. Conversely, an excessively large adhesive region can compromise the flexibility of the aggregates, resulting into a breakup of the aggregates.}
\end{figure}

The results reported in the previous two sections rely on the assumption that the adhesive portion of the sperm head is localized at the forward tip. In order to study how the structure of the adhesive region affects the performance of the aggregate we have simulated five variants with aspect ratio $b/a =0.5$ and $c/b = 0.4$, which is close to features observed in {\em P. maniculatus} and {\em P. polionotus} \cite{Fisher:2014}. We define the adhesive region as a strip that starts from a position $\Delta x$ from the tip of the sperm head and extends toward the tip with a thickness $\Delta t$, i.e. when $\Delta t = \Delta x$ then the front of the sperm cell is covered in an adhesive cap that extends down to $\Delta x$, whereas if $\Delta t < \Delta x$ the adhesive region is limited to a band (Fig.~\ref{fig:AcroPhase}a).

Fig. \ref{fig:AcroPhase}b shows the typical velocity increase obtained from the simulations of the different variants. The best performance is obtained when the adesive strip occupies a small portion of the sperm head. On the other hand, when the adhesive strip extends towards the tip, i.e. $\Delta t \sim \Delta x$, the advantage is somewhat lost. If the adhesive strip extends past the mid-region of the cell toward the tail, we see that the aggregates become unstable and break up over time, denoted by the black region on the right side of Fig.~\ref{fig:AcroPhase}b. 

This behavior can be understood by considering the torque exerted on the sperm cells by the adhesion molecules. The adhesion molecules pull the adhesive region of neighboring cells together. Since the cells cannot overlap, the point of contact between the cells acts like a fulcrum applying a torque to the sperm cell. When the adhesion molecule is toward the front of the sperm head, it will generally cause the orientations of the two cells to converge. As explained earlier, this can cause a slowdown of the group, but ultimately promotes adhesion as the resulting configuration has the particles swimming toward each other. If the adhesive region is toward the tail of the cell, the effect of the torque is for the velocities to diverge, resulting in a configuration in which the cells are swimming away from each other and the aggregate eventually breaks up. This leads to the unstable region shaded black on Fig.~\ref{fig:AcroPhase}d. Such an effect is amplified in wider cells as the point of contact is generally further from the centre line of the cell. The cells are generally fastest when the adhesion region is limited to a thin band round the cell, i.e. $\Delta t \sim 0.1$. This is the configuration which generally minimises the torque on the velocity of the cell, when the cells are parallel, leading  to the fastest aggregates.

Although the breakup of the aggregate due to the torque build-up is likely a property of the ellipsoidal head shape considered here, our results suggest that the an efficient aggregation might require the cells to maintain some freedom to rotate in order to guarantee some degree of flexibility of the whole aggregate. 

\subsection{Aggregation performance of {\em P. maniculatus} from morphometric data}

\begin{figure}[t]
\includegraphics[width = \columnwidth]{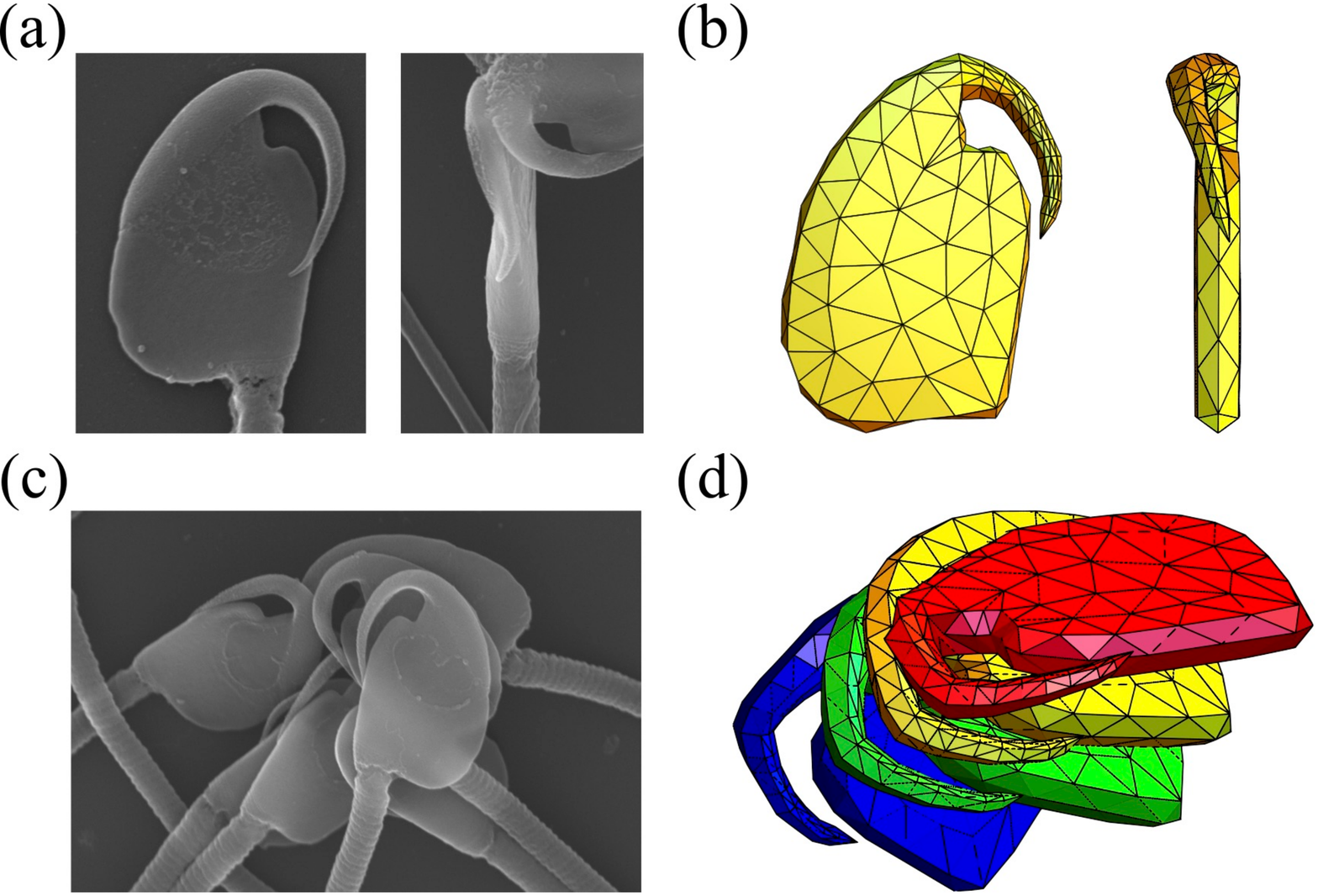}
\caption{\label{fig:3D_Model} (a) Electron micrographs of sperm from {\em P. maniculatus}, a species in which sperm aggregate \cite{Fisher:2014}. (b) Planar views of the sperm head model based constructed from the morphometric data. (c) Electron micrograph of an aggregate of {\em P. maniculatus} sperm cells. (d) A snapshot from a simulation of a group of six sperm heads using a more realistic model head shape.}
\end{figure}

In this section, we extend our approach to account for more realistic head geometries. In particular, we consider the sperm cell of {\em P. maniculatus}. The purpose of this analysis is two-fold. First, using realistic head geometries allows us to test the validity of the results presented in the previous sections and to demonstrate that many of the properties illustrated with the elliptical model are robust and carry over to realistic head geometries. Second, we explore the interplay between the geometry of the apical hook and the structure of the adhesive region and identify a number of highly efficient configurations, that could provide a basis for future experimental studies. 

Sperm samples were dissected from the epididymes of sexually mature males, fixed in 2.5\% glutaraldehyde, dehydrated first in a graded ethanol series, then in hexamethyldisilazene, and sputter coated in gold/palladium. Samples were imaged on a Hitachi SU-3500 scanning electron microscope at the Laboratory for Biological Ultrastructure, University of Maryland (Fig.~\ref{fig:3D_Model}a,c). The two planar electron micrographs shown in Fig.~\ref{fig:3D_Model}a,c were first traced to create a general outline and identify the center-line of the hook. This was then extruded to a 3D model consisting of 400 triangles (Fig.~\ref{fig:3D_Model}b). 

\begin{figure}[t]
\includegraphics[width = \columnwidth]{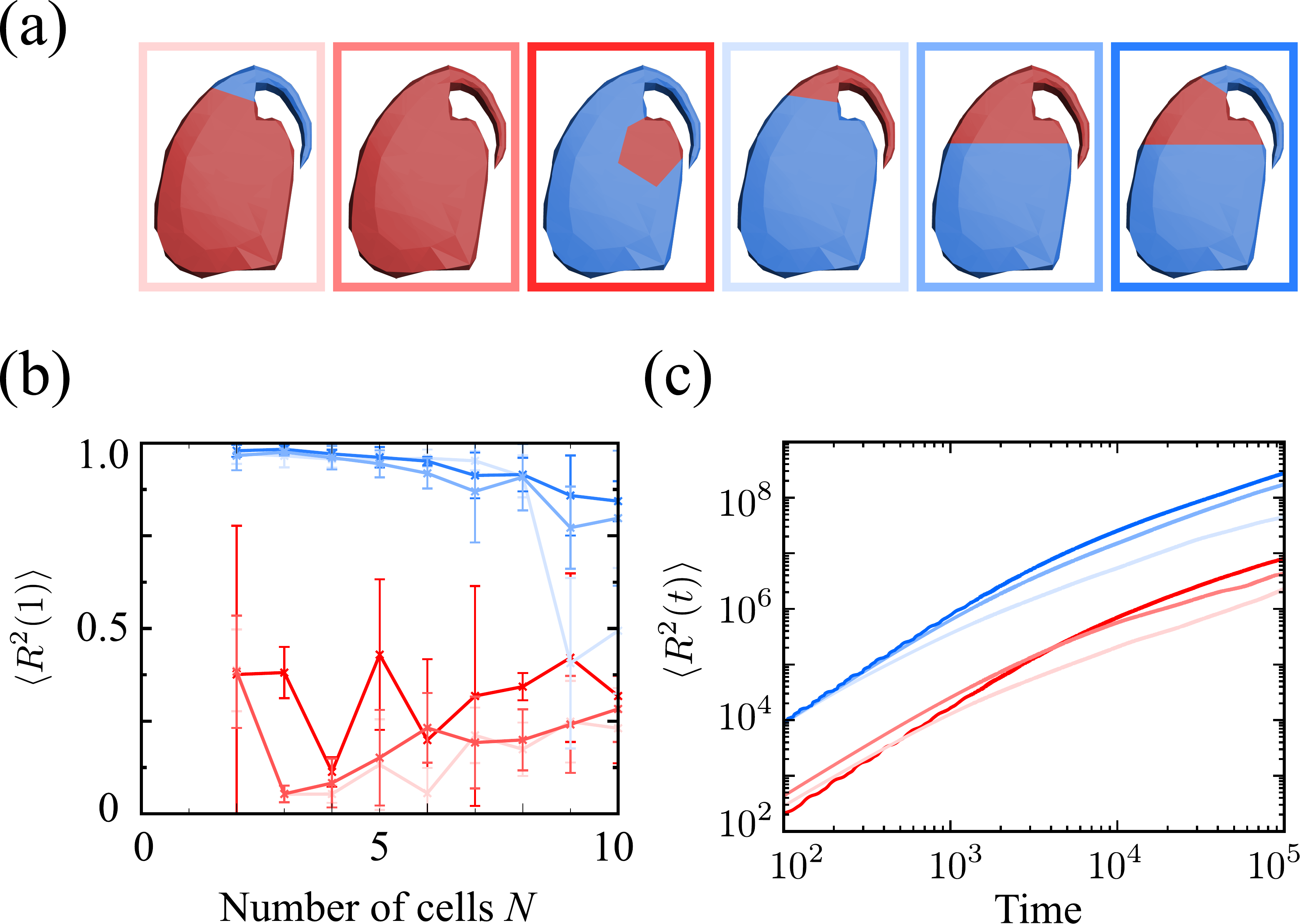}
\caption{\label{fig:model_results} (a) Instantaneous speed of sperm aggregates depends strongly on the position of the adhesive region on the sperm head. (d) Mean square displacement of a group of 4 sperm with the accurate head shape. Again, the position of the adhesive region greatly affects the efficiency of the aggregate.}
\end{figure}

Sperm aggregates were simulated in small groups such as that imaged in Fig.~\ref{fig:3D_Model}d. To shed light on the interplay between the hook geometry and the structure of the adhesive region, we considered six different variants as shown in Fig.~\ref{fig:model_results}a. The instantaneous speed of the aggregates (Fig. ~\ref{fig:model_results}b) is particularly sensitive to the structure of the adhesive region in the presence of the hook. Aggregates with the adhesive region covering only the mid-section or hook are able to maintain consistently high instantaneous speeds, whereas sperm cells where the adhesive region extends beyond the midpoint of the head are significantly slower  (Fig.~\ref{fig:model_results}b). At longer time scales the difference between the aggregates is even more pronounced, with sperm cells having an adhesive cap on the forward section of the head being the fastest and sperm cells with the whole body adhesive being the slowest (Fig.~\ref{fig:model_results}c). These behaviors are rooted in the physical mechanism outlined in Sec. \ref{sec:results}C. The torques applied by the adhesion molecules are not necessarily promoting alignment unless limited to the upper half of the sperm head. The {\em P. maniculatus} sperm head is substantially flat, with two large roughly parallel sides which reduce the torque exerted by adjacent sperm allowing for very efficient stacking. This cell shape reduces the crowding effect and desegregation observed for the ellipsoidal sperm heads, resulting in generally more robust and efficient sperm aggregates.


\section{\label{sec:discussion}Discussion and conclusions}

Using numerical simulations, we have explored how the geometry of the sperm head affects the performance of aggregation. In our simulations, the sperm head is modeled as a persistent random walker characterized by a non-trivial three-dimensional shape. The head is further equipped with an adhesive region, where the cell binds to the head of other cells. Upon modeling the head via a simple parametric shape (i.e. an ellipsoid), we have established that the mobility of sperm aggregates is strongly affected by the head geometry. In particular, slender and oblate head shapes lead to significantly faster aggregates compared to spherical head shapes, because they allow the cells to form tightly  packed and highly polarized aggregates, thus reducing the spread in the velocities of the individual cells.  

The structure of the adhesive region, also has a profound impact on both the stability and the swimming performance of the aggregates. In particular, by simulating ellipsoidal heads endowed of adhesive strips of varying thickness and location, we find that some degree of internal flexibility is a vital feature for successful aggregation. When aggregates first form they typically do not have a polar structure and the cells tend to swim against each other. Cells having large adhesive areas, are too tightly bound to their neighbours to reorient and eventually break apart. On the other hand, cells with smaller adhesive areas, such as a strip in the vicinity of the head equatorial plane, are sufficiently mobile to reorient in such a way that all the cells move toward the same direction. This enhances the stability of the aggregates as well as the persistence of their random motion. 

Finally, most muriod rodents produce sperm with one or more hooks on the head of the cell \cite{Breed:2005}. Although the function of sperm hooks is unresolved, it has been hypothesized that the hook may interact with the epithelium of the female oviduct to aid sperm migration \cite{Suarez:1987}, may offer a hydrodynamic advantage to singly swimming cells \cite{Montoto:2011}, and/or may permits cells to attach to the hooks or flagellum of other sperm cells, thereby creating aggregates \cite{Immler:2007} (but see Ref. \cite{Tourmente:2016}). In this latter prediction, evidence suggests that sperm-sperm adhesion occurs along the inner surface of the hook wood mice ({\em Apodemus sylvaticus}) \cite{Moore:2002}. Our model suggests another alternative, that the hook acts as a shield in combination with an adhesive region localized at the acrosome or directly below at the equatorial segment. By shielding one side of the sperm head, the hook could favor the formation of aggregates in which all the cells have the same orientation, leading to more dense packing and a more effective alignment of the velocities of the interacting cells. This hypothesis is consistent with empirical observations of {\em Peromsycus} sperm \cite{Fisher:2014} and numerical simulations (Fig. \ref{fig:model_results}). 

\begin{acknowledgments}
This work was supported by the Netherlands Organization for Scientific Research (NWO/OCW), as part of the Frontiers of Nanoscience program and the Vidi scheme (DJGP, LH, LG), the National Institutes of Health (NICHD K99/R00 HD071872) (HSF) and the National Science Foundation (1711817) (KAH).
\end{acknowledgments}

\end{document}